\documentclass[11pt]{article}
\usepackage{amsmath,amssymb,color,graphics,epsfig,cite}
\usepackage{graphicx,subfigure}
\usepackage{amsfonts}
\newcommand{\hoch}[1]{$\, ^{#1}$}

\newcommand{\be}{\begin{equation}}
	\newcommand{\ee}{\end{equation}}
\newcommand{\bea}{\setlength\arraycolsep{2pt} \begin{eqnarray}}
	\newcommand{\eea}{\end{eqnarray}}
\newcommand{\nn}{\nonumber}

\def\fft#1#2{{\frac{#1}{#2}}}

\def\0{{\sst{(0)}}}
\def\1{{\sst{(1)}}}
\def\2{{\sst{(2)}}}
\def\3{{\sst{(3)}}}
\def\4{{\sst{(4)}}}
\def\5{{\sst{(5)}}}
\def\6{{\sst{(6)}}}
\def\7{{\sst{(7)}}}
\def\8{{\sst{(8)}}}
\def\sst#1{{\scriptscriptstyle #1}}

\begin{document}

\begin{center}

{\Large {\bf Quadratic Curvature Correction to the Euclidean Action of Rotating AdS Black Holes in General Dimensions}}
		
\vspace{20pt}
		
Si-Yue Lu\hoch{1},  H. L\"{u}\hoch{2,1}
		
\vspace{10pt}

{\it \hoch{1}The International Joint Institute of Tianjin University, Fuzhou,\\ Tianjin University, Tianjin 300072, China}
\medskip

{\it \hoch{2}Center for Joint Quantum Studies, Department of Physics,\\
			School of Science, Tianjin University, Tianjin 300350, China }
%\medskip

%{\it \hoch{3}Peng Huanwu Center for Fundamental Theory, Hefei, Anhui 230026, China}
		
\vspace{40pt}

\underline{ABSTRACT}
\end{center}

We adopt the improved Reall-Santos method to obtain the leading-order perturbative correction of the quadratic curvature invariants to the on-shell Euclidean action of rotating anti-de Sitter (AdS) black holes in general $D$ dimensions. The corresponding Gibbs free energy is a function of thermodynamic variables, temperature and angular velocities, which are unperturbed in this approach.

\vfill{siyue\_lu@tju.edu.cn\ \ \ mrhonglu@gmail.com}
	
\thispagestyle{empty}
\pagebreak
	
%\voffset=-40pt
%\setcounter{page}{1}
	
%\tableofcontents
%\addtocontents{toc}{\protect\setcounter{tocdepth}{2}}
%\newpage

\section{Introduction}
\label{sec:introduction}

The pursuit of a consistent theory of quantum gravity remains one of the central challenges in theoretical physics. In the absence of a consistent ultraviolet (UV) completion, the effective field theory (EFT) method in a perturbative approach provides a powerful framework for describing the quantum effects in low-energy gravitational phenomena. Within this paradigm, the Einstein-Hilbert action is regarded as the leading-order term, with higher-derivative interactions, suppressed by the UV cutoff scale, emerging as inevitable corrections from integrating out massive degrees of freedom. Among these, the quadratic curvature terms \(R^2\), \(R^{\mu\nu} R_{\mu\nu}\), and \(R^{\mu\nu\rho\sigma} R_{\mu\nu\rho\sigma}\) constitute the most pertinent next-to-leading order corrections, capturing the first glimpses of physics beyond classical General Relativity. (This perturbative EFT approach is different from Einstein gravity extended classically with quadratic curvature invariants which can lead to renormalizability at the price of introducing ghostlike massive graviton modes \cite{Stelle:1976gc,Stelle:1977ry}.) Black hole thermodynamics provides a crucial testing ground for such corrections, but traditional methods face severe limitations. The need to solve perturbed field equations order-by-order and construct intricate boundary terms makes the study of thermodynamic properties prohibitively difficult in generic settings, beyond spherically-symmetric and static spacetimes.

	A major advance came from Reall and Santos (RS) \cite{Reall:2019sah}, who showed that for asymptotically flat Kerr black holes, thermodynamic corrections could be derived without solving the perturbed field equations by working in the grand canonical ensemble via the Euclidean action approach based on the quantum statistic relation (QSR) \cite{Gibbons:1976ue}. The RS method achieves two simplifications. One is to remove the need to solve for the perturbative solution, which can be a significant task for rotating black holes in general dimensions. The other, for asymptotically flat black holes, is to remove the need to know the structures of the Gibbons-Hawking-York (GHY) terms \cite{Gibbons:1976ue,York:1972sj} associated with the higher-derivative corrections in the bulk action. This method was adopted to obtain the quadratic curvature corrections \cite{Wu:2024iiz} to the Myers-Perry (MP) black holes \cite{Myers:1986un} in general dimensions. The RS method was also generalized to include next-to-next-to-leading order corrections and beyond \cite{Ma:2023qqj}. (It is worth noting that for string or some string-inspired theories, the complete set of thermodynamic quantities can be obtained directly from the theory, without having to know the black hole solutions \cite{Lu:2025eub,Yang:2025rud}.)

The situation becomes subtler when a cosmological constant is involved. For AdS black holes, the quadratic curvature correction may not decay asymptotically fast enough since it could alter the AdS radius itself. Consequently, the boundary terms, such as the GHY term and the holographic counterterms, are generally needed and are difficult to derive for generic higher-derivative corrections in the bulk, as demonstrated with the Gauss-Bonnet bulk term in \cite{Reall:2019sah}. Two approaches were developed to avoid these cumbersome calculations. One is the background subtraction method of Hawking and Page \cite{Hawking:1982dh}. Choosing an appropriate background for subtraction becomes the key, and the work was primarily focused on AdS black holes with spherical symmetry so that the result can be tested easily by a perturbative method \cite{Xiao:2023two,Xiao:2025icr}. There has been a recent progress in understanding the background subtraction and the quadratic curvature correction to the Euclidean action of the $D=5$ rotating AdS black hole \cite{Hawking:1998kw} was correctly re-derived by the background subtraction method \cite{Guo:2025ohn,Chen:2025ary}.

Another more effective approach is to make use of the fact that in perturbative EFT, the theories are equivalent under field redefinition up to the corresponding order \cite{Hu:2023gru,Ma:2024ynp}. In pure cosmological gravity we consider in this paper, we can perform field redefinition on the metric by curvature tensors so that Einstein gravity with generic quadratic curvature corrections can be transformed to Einstein gravity with the Weyl-squared term. For AdS black holes, the Weyl-squared term decays so rapidly at the asymptotic region that the effective cosmological constant is unmodified; it is the same as the bare cosmological constant. The corresponding on-shell boundary GHY term and holographic counterterms can be ignored in the on-shell Euclidean action \cite{Hu:2023gru}. We can thus apply the RS method as if it were the asymptotically flat case. After obtaining the Weyl-squared correction to the black hole thermodynamics, we can reverse the field redefinition and obtain the generic corrections. Note that the same field redefinition can be applied to different black holes in the theory. This improved RS method for AdS black holes thus completely restores the computational efficiency of the original RS method. Its validity was explicitly confirmed for the first time by the perturbative-solution method for rotating AdS black holes with two equal angular momenta in five dimensions \cite{Ma:2024ynp}.

	In this paper, we shall apply this improved RS method and derive the generic quadratic curvature corrections to black hole thermodynamics of the general rotating AdS black holes constructed in \cite{Gibbons:2004uw,Gibbons:2004js}. In section 2, we review the field redefinition that related Einstein-Weyl gravity with a bare cosmological constant to gravity with generic quadratic curvature corrections. We then obtain the general corrections to the on-shell Euclidean action of the AdS rotating black holes, and also present a few specific examples in section 3.  We conclude the paper in section 4.

\section{Einstein gravity with quadratic curvature corrections}

The leading-order correction to cosmological Einstein gravity in general $D$ dimensions involves three quadratic Riemann-tensor invariants. The total bulk Lagrangian up to this order is
\be
{\cal L}=\sqrt{-g}\Big( \sigma_0 \big(R - 2\Lambda_0\big) + \delta L\Big)\,,\qquad \delta L= \sigma_0^2 \big(c_1 R^2 + c_2 R^{\mu\nu} R_{\mu\nu} + c_3 R^{\mu\nu\rho\sigma} R_{\mu\nu\rho\sigma}\big)\,.\label{c123lag}
\ee
Here we introduce a bookkeeping order parameter $\sigma_0$, which we can set to identity after all the dust settles. For our purpose, we consider a negative bare cosmological constant
\be
\Lambda_0 =-\fft{(D-1)(D-2)}{2\ell_0^2}\,,
\ee
where $\ell_0$ is the AdS radius. We also define $g_0=1/\ell_0$, typically used in supergravity convention, which should not be confused with the metric determinant. The Euclidean on-shell action requires further knowledge of the boundary terms such as the GHY term and holographic counterterms \cite{Balasubramanian:1999re,Bianchi:2001kw}. For the Gauss-Bonnet combination, the holographic counterterms were constructed in \cite{Brihaye:2008xu,Liu:2008zf}. For general quadratic curvature gravity with generic coupling constants $(c_1,c_2,c_3)$, a method of deriving these surface terms was proposed in \cite{Cremonini:2009ih}. The method is too involved if the purpose is only to derive the on-shell action for rotating AdS black holes in general dimensions.

It is understood that the higher-order corrections in \eqref{c123lag} are perturbative with small coupling constants $(c_1,c_2,c_3)$. Up to this order, we consider field redefinition
\be
g_{\mu\nu}\ \rightarrow\  g'_{\mu\nu}=g_{\mu\nu}+\lambda_0\, g_{\mu\nu}+\sigma_0\,\lambda_1\, R_{\mu\nu}+\sigma_0\, \lambda_2\, g_{\mu\nu}R\,,
\ee
where the parameters are chosen as \cite{Hu:2023gru,Ma:2024ynp}
\bea
\ell_0 &\rightarrow& \ell_0'=\ell_0+\frac{\ell_0}{2}
\Big(\lambda_0-\sigma_0\frac{(D-1)(D-2)(\lambda_1+D\lambda_2)}{2\ell_0^2}\Big)
\,,\nn\\ \sigma_0&\rightarrow&\sigma_0'=\sigma_0-\fft{\sigma_0}2(D-2)\Big(
\lambda_0+\sigma_0 \frac{(D-1)(\lambda_1+D\lambda_2)}{\ell_0^2}\Big)\,,\nn\\
\lambda_0&=&(D-1)\sigma_0(\lambda_1+D\lambda_2)\ell_0^{-2}\,,\qquad
\lambda_1=-c_2-\frac{4c_3}{D-2} \,,\nn\\
\lambda_2 &=& \frac{4c_3+2(D-1)c_1+(D-1)c_2}{(D-1)(D-2)}\,,\label{couplingredef}
\eea
the Lagrangian \eqref{c123lag} then transforms perturbatively to Einstein-Weyl gravity
\be
{\cal L}=\sqrt{-g}\Big( \sigma_0 \big(R - 2\Lambda_0\big) + \sigma_0^2\,
\alpha C^{\mu\nu\rho\sigma} C_{\mu\nu\rho\sigma}\Big)\,,\label{EWlag}
\ee
where we have renamed the coupling constant $c_3$ as $\alpha$. This should not be confused with non-perturbative critical gravity \cite{Lu:2011zk}. The need to transform the bare cosmological constant makes the field redefinition more complicated, compared to the case without the cosmological constant.

For a typical AdS black hole with sphere topology, the Weyl tensor decays so rapidly that we do not need to be concerned with its boundary terms at the asymptotic infinity. We can thus apply the RS method directly and its correction to the on-shell Euclidean action is simply given by the bulk action alone:
\be
\delta I=\frac{\sigma_0^2}{16\pi}\int_{\mathcal{M}}\sqrt{-g}\, \alpha C^{\mu \nu\rho\sigma} C_{\mu \nu\rho\sigma}\,,\label{EWcor}
\ee
which is evaluated on the original unperturbed black hole solution.  We can read off the Weyl-squared correction to the Gibbs free energy $\delta G_{\rm Weyl}$. Reversing the redefinition \eqref{couplingredef}, we obtain the general quadratic curvature corrections to the Gibbs free energy $\delta G_{c_i}$. The corresponding corrections to the on-shell Euclidean action becomes
\be
\fft{\delta G_{c_i}}{T_0}=\delta I\equiv\frac{1}{16\pi}\int_{\mathcal{M}}\sqrt{-g}\, \delta L +
\int_{\partial {\cal M}} {\rm boundary\ terms}\,,\label{gencor1}
\ee
without having to know the exact boundary terms, where $\delta L$ is given by \eqref{c123lag}. For later purpose, it is more convenient to recombine the three quadratic curvature invariants by defining
\be
c_1=\frac{2\alpha}{(D-1)(D-2)}-\frac{\beta}{D}+\gamma\,,\quad c_2=-\frac{4\alpha}{D-2}+\beta\,,\quad c_3=\alpha\,,
\ee
so that the $\delta L$ in \eqref{c123lag} becomes
\be
\delta L=\sigma_0^2\big(\alpha\, C^{\mu\nu\rho\sigma}C_{\mu\nu\rho\sigma}+\beta\, r^{\mu\nu}r_{\mu\nu}+\gamma\, R^2\big)\,,\label{abclag}
\ee
where $r_{\mu\nu}=R_{\mu\nu}-\frac{1}{D}g_{\mu\nu}R$, which vanishes for any Einstein metrics.

\section{Corrections to general rotating AdS black holes}

\subsection{Rotating AdS black holes in general $D$ dimensions}

In this section, we apply the improved RS method to compute the leading quadratic curvature corrections to black hole thermodynamics of rotating AdS black holes in general $D$ dimensions. The solutions were constructed in \cite{Gibbons:2004uw,Gibbons:2004js}. The Einstein metric with $\ell_0$ AdS radius is
\bea
ds^2 &=& -W(1+r^2\ell_0^{-2})d\tau^2+\frac{2m}{U}\Big(Wd\tau - \sum_{i=1}^{N}\frac{a_i\mu_i^2d\phi_i}{\Xi_i}\Big)^2+
\sum_{i=1}^{N}\frac{r^2+a_i^2}{\Xi_i}\mu_i^2d\phi_i^2 \nn\\
&& +\frac{Udr^2}{V-2m}+\sum_{i=1}^{N+\epsilon}\frac{r^2+a_i^2}{\Xi_i}
d\mu_i^2-\frac{\ell_0^{-2}}{W(1+r^2\ell_0^{-2})}\Big(\sum_{i=1}^{N+\epsilon}
\frac{r^2+a_i^2}{\Xi_i^2}\mu_id\mu_i\Big)^2\,,\label{genadsrot}
\eea
where
\bea
W= \sum_{i=1}^{N+\epsilon}\frac{\mu_i^2}{\Xi_i}\,, \qquad U=r^\epsilon\sum_{i=1}^{N+\epsilon}\frac{\mu_i^2}{r^2+a_i^2}
\prod_{j=1}^{N}(r^2+a_j^2)\,,\nn\\
V=r^{\epsilon-2}(1+r^2\ell_0^{-2})\prod_{i=1}^{N}(r^2+a_i^2)\,,\qquad \Xi_i = 1 -a_i^2\ell_0^{-2}\,,
\eea
and $\epsilon$ here is a discrete number given by $D=2N+1+\epsilon$. For odd $D$, we have $\epsilon=0$ and for even $D$, we have $\epsilon=1$. The solution contains $N+1$ integration constants, $m$ and $a_i$, $i=1,2,\ldots,N$. They parameterize the mass and $N$ independent orthogonal rotations. The thermodynamic properties were analysed in \cite{Gibbons:2004ai}, and here is the list of complete thermodynamic quantities
\bea
M_0&=&\left\{\begin{array}{lc}
	\frac{\sigma_0 m \mathcal{A}_{D-2}}{4\pi (\prod_j
		\Xi_j)}(\sum^N_{i=1}\frac{1}{\Xi_i}-\frac{1}{2})\,,&\quad \hbox{$\epsilon=0$ ,} \\
	\frac{\sigma_0 m \mathcal{A}_{D-2}}{4\pi (\prod_j
		\Xi_j)}\sum^N_{i=1}\frac{1}{\Xi_i}\,, & \quad\hbox{$\epsilon=1$ ,}
\end{array}\right.\nn\\
&&\nn\\
S_0&=&\left\{\begin{array}{lc}
	\frac{\sigma_0 \mathcal{A}_{D-2}}{4r_+}\prod_i\frac{r_+^2+a_i^2}{\Xi_i}\,,&\quad \hbox{$\epsilon=0$ ,} \\
	\fft{\sigma_0 \mathcal{A}_{D-2}}4\prod_i\frac{r_+^2+a_i^2}{\Xi_i}\,, &\quad\hbox{$\epsilon=1$ ,}
\end{array}\right.\qquad\qquad\qquad\qquad
J_{0i} =\frac{\sigma_0 m a_i \mathcal{A}_{D-2} }{4 \pi \Xi_i (\prod_j \Xi_j)}\,,\nn\\
&&\nn\\
T_0&=&\left\{\begin{array}{lc}
	\frac{r_+(1+g_0^2 r_+^2)}{2\pi }\sum^N_{i=1}\frac{1}{r_+^2+a_i^2}-\frac{1}{2\pi r_+}\,,&\quad \hbox{$\epsilon=0$ ,} \\
	\frac{r_+(1+g_0^2 r_+^2)}{2\pi }\sum^N_{i=1}\frac{1}{r_+^2+a_i^2}-\frac{1-g_0^2 r_+^2}{4\pi r_+}\,, & \quad\hbox{$\epsilon=1$ ,}\\
\end{array}\right.\qquad
\Omega_{0i}=\frac{(1+g_0^2 r_+^2)a_i}{r_+^2+a_i^2}\,,\label{unperturbedthermo}
\eea
where $r_+$ is the largest root of $V(r)-2m=0$ and $\mathcal{A}_{D-2}$ is the volume of the unit $(D-2)$ sphere, i.e.~$\mathcal{A}_{D-2}=2\pi^{(D-1)/2}/\Gamma[(D-1)/2]$. Note again that we used $g_0=1/\ell_0$ to simplify the expression. It is easy to verify that the first law of the black hole thermodynamics holds, namely
\be
dM_0 = T dS_0 + \sum_i\Omega_{0i}\, dJ_{0i}\,.
\ee
The Gibbs free energy $G_0$ associated with the on-shell Euclidean action is related to the mass of the black hole via the Legendre transformation, namely
\be
G_0=\fft{\sigma_0 {\cal A}_{D-2} (1-g_0^2 r_+^2) m}{8\pi (\prod_i \Xi_i) (1+g_0^2 r_+^2) } = M_0 - T_0 S_0 - \sum_i\Omega_{0i} J_{0i}\,.\label{G0res}
\ee
Thus, the Gibbs free energy corresponds to an ensemble with temperature and angular velocities as it basic thermodynamic variables. The Euclidean action associated with different ensembles based on Legendre transformations in the Kaluza-Klein approach was recently proposed \cite{Ma:2026cej}. It is of interest to note that the mass $M_0$ and angular momenta $J_{0i}$ satisfy the following relation
\be
M_0=\left\{
    \begin{array}{ll}
      \Sigma_{i=1}^N \fft{J_{0i}}{a_i}(1-\fft{\Xi_i}{2N})\,, &\qquad \epsilon=0 \\
       & \\
      \Sigma_{i=1}^N \fft{J_{0i}}{a_i}\,,&\qquad \epsilon=1
    \end{array}
  \right.
\ee

\subsection{The Weyl-squared correction}

There is a great simplification when we compute the Weyl-squared term for the rotating AdS black holes in general dimensions. Compared to the MP metric, the Einstein metric \eqref{genadsrot} of rotating AdS black holes is significantly more complicated. However, when we substitute the solution into $C^{\mu \nu\rho\sigma}C_{\mu \nu\rho\sigma}$, we find that the cosmological constant, associated with $\ell_0$ drops out completely, as if it were the Ricci-flat metric. The contribution of the cosmological constant enters the Euclidean action only via the determinant of the metric as an overall coefficient
\be
\sqrt{-g} = (\prod_i^N \Xi_i)^{-1}\, \Big(\sqrt{-g}\Big|_{\Lambda_0=0}\Big)\,.
\ee
The origin of this overall coefficient is due to the fact that the $N$ azimuthal angles $\phi_i$  should be scaled by a constant $\Xi_i$ to maintain the 2$\pi$ period when the solution is promoted from the MP solution to the rotating AdS black holes. Thus the Weyl-squared correction to the Gibbs free energy of the rotating AdS black holes is simply given by
\be
\delta G_{\rm AdS-Weyl}=(\prod_i^N \Xi_i)^{-1}\,\delta G_{\rm MP-Weyl}\,,\label{Galpha}
\ee
where the Weyl-squared correction to the Gibbs free energy of the general MP black holes is known in literature, given by \cite{Wu:2024iiz}
\bea
\delta G_{\rm MP-Weyl}&=&\left\{\begin{array}{lc}
	(-1)^N\frac{\sigma_0^2 \mathcal{A}_{D-2}\alpha(D-3)m^2}{4\pi}\sum_{i}^{N}
\frac{r_+^4-6a_i^2r_+^2+a_i^4}{b_i(r_+^2+a_i^2)^3}\,,  &\qquad \hbox{$\epsilon=0$ ,} \\
&\\
	(-1)^N\frac{\sigma_0^2 \mathcal{A}_{D-2}\alpha(D-3)m^2}{\pi}\sum_{i}^{N}\frac{r_+(r_+^2-a_i^2)}{b_i(r_+^2+a_i^2)^3}
\,,& \qquad\hbox{$\epsilon=1$ ,}
\end{array}\right.\label{mpres}
\eea
where $b_i=\prod_{j\ne i} (a_i^2-a_j^2) $. Note that for the Ricci-flat MP black holes, the Weyl-squared term is equivalent to the Riemann-squared term. It should be pointed out that in the above expressions, although there is no explicit dependence on the cosmological constant, there is implicit dependence since $r_+$ and $m$ are related by $V(r_+)=2m$, which depends on the cosmological constant. The total corrected Gibbs free energy for the rotating AdS black hole in Einstein-Weyl gravity with a cosmological constant is
\be
G_{\rm EW-AdS} = G_0 + (\prod_i^N \Xi_i)^{-1}\,\delta G_{\rm MP-Weyl}\,,
\ee
where the first and the second terms above are given by \eqref{G0res} and \eqref{mpres} respectively.

\subsection{General quadratic curvature corrections}

Having obtained the Weyl-squared correction, we are now in the position to derive the general quadratic curvature corrections, by reversing the field redefinition. There is however a subtlety that should first be addressed. The thermodynamic variables under the Gibbs free energy, namely the temperature and angular velocities of the uncorrected rotating AdS black holes \eqref{unperturbedthermo} depend on $\ell_0=1/g_0$ and $\sigma_0$, which both transform under the field redefinition. Thus, these thermodynamic variables transform under the field redefinition. However, it is clearly more convenient to fix these variables in the ensemble of the Gibbs free energy. We thus perform a reparametrization on the $(N+1)$ integration constants $(r_+,a_i)$ to $(r_+ + \delta r_+, a_i + \delta a_i)$ so that after field redefinition, the temperature and angular velocities remain uncorrected.  This condition requires
\be
\delta a_i = -\frac{a_i r_+ \big(4 \Xi_i\delta r_+ -(D-4) \lambda_0\,g_0^2 r_+ \left(a_i^2+r_+^2\right)\big)}{2 \left(1+g_0^2 r_+^2+1\right) \left(r_+^2-a_i^2\right)}\,,
\ee
for both odd or even dimensions, together with
\be
\delta r=
\left\{
  \begin{array}{ll}
-\frac{(D-4)\lambda_0\, g_0^4 r_+^5\,\zeta}{2 \left(
1- g_0^3 r_+^3\,\eta + g_0^2 r_+^2 (1+3 g_0^2 r_+^2)\,\zeta\right) }\,, & \qquad\epsilon=0\,, \\
    & \\
-\frac{(D-4)\lambda _0\, g_0^2  r_+^3 \left(1+2g_0^2 r_+^2\,\zeta\right)}{2 \left(1+ g_0^2 r_+^2 -2 g_0^3 r_+^3\, \eta + 2 g_0^2 r_+^2 (1 + 3g_0^2 r_+^2)\, \zeta\right)}\,, &\qquad
\epsilon=1\,,
  \end{array}
\right.
\ee
where the dimensionless parameter $(\zeta,\eta)$ are
\be
\zeta =\fft{1}{g_0^2} \sum_{i=1}^N \fft{1}{r_+^2+a_i^2}\,,\qquad
\eta = \fft{2r_+}{g_0^3} \sum_{i=1}^N \fft{1 + g_0^2 (r_+^2 - 2 a_i^2)}{
r_+^4 - a_i^4}\,.
\ee
Performing the field redefinition, and also the integration constant reparametrization, we obtain the general corrected Gibbs free energy associated with the on-shell Euclidean action of \eqref{abclag}, given by
\be
G= G_0+\delta G_{\alpha}+\delta G_{\beta} +\delta G_{\gamma}\,,
\ee
where $\delta G_{\alpha}$ is simply the one given in \eqref{Galpha} and we shall not repeat here. The $\beta$ correction term $\Delta G_{\beta}$ vanishes identically, since for Einstein metrics, the tensor $r_{\mu\nu}=0$. The $\delta G_\gamma$ associated with the $R^2$ term is more involved, given by
\bea
\delta G_{\gamma}&=&\left\{\begin{array}{lc}
\frac{\gamma \sigma_0^2 D(D-1)\mathcal{A}_{D-2}g_0^2}{16 \pi (D-2) r_+^2}\frac{\prod^N_{i=1}\rho_i^2}{\Theta_N^2}[(\frac{1}{2}(D^2-3D+4)\Theta_N+(D-4)
\Theta_{N-1})g_0^2 r_+^2 & \nn\\
-\frac{1}{2}(D^2-D-4)\Theta_N+(D-4)\Theta_{N-1}]\,, &\hbox{$\epsilon=0$ ,}\\
&\\
\frac{\gamma \sigma_0^2 D(D-1)\mathcal{A}_{D-2}g_0^2}{16 \pi (D-2) r_+}\frac{\prod^N_{i=1}\rho_i^2}{\Theta_N^2}[(\frac{1}{2}D(D-2)
\Theta_N+(D-4)\Theta_{N-1})g_0 r_+^2 &\nn\\
-\frac{1}{2}D(D-2)\Theta_N+(D-4)\Theta_{N-1}]\,, &\hbox{$\epsilon=1$ ,}
\end{array}\right.
\eea
where
\bea
\Theta_k=\frac{1}{k!}\frac{d^k}{d\epsilon^k}\prod^N_{i=1}(1+\epsilon \Xi_i)\,,\qquad\rho_i^2=r_+^2+a_i^2\,.
\eea
Note that $\delta G_\gamma$ has an overall $g_0^2$ coefficient, which vanishes for asymptotic-flat MP black holes that are Ricci flat.

\subsection{Some explicit examples}

Having obtained the quadratic curvature correction to Gibbs free energy of the rotating AdS black holes in general $D$ dimensions via the Euclidean action, we note that the general formulae are complicated. We thus present a few simpler explicit examples, and compare the results in literature.

\subsubsection{$D=4$ and $D=5$}

We first consider $D=4$, following from the general formula, the corrected Gibbs free energy is
\bea
G_4 &=& \frac{\sigma _0 \left(r_+^2+a^2\right) \left(1-g_0^2 r_+^2\right)}{4 \Xi_a\, r_+}\nn\\
&& - \fft{\sigma_0^2}{\Xi_a\,r_+} \Big(\frac{\alpha  \left(r_+^2-a^2\right) \left(1+g_0^2 r_+^2\right)^2}{\left(r_+^2+a^2\right)}+6 \gamma  g_0^2 \left(r_+^2+a^2\right) \left(1-g_0^2 r_+^2\right)\Big),
\eea
where $\Xi_a=1-a^2 g_0^2$. The temperature and angular velocity, by the construction, are uncorrected, given by
\be
T=T_0=\frac{3 g_0^2 r_+^4+r_+^2 -a^2 \left(1-g_0^2 r_+^2\right)}{4 \pi  r_+ \left(r_+^2+a^2\right)}\,,\qquad \Omega=\Omega_0=\frac{a \left(1+g_0^2 r_+^2\right)}{r_+^2+a^2}\,.
\ee
The remaining thermodynamic quantities can be obtained, given by
\bea
S &=&-\fft{\partial G}{\partial T}\Big|_\Omega= \fft{\sigma_0}{\Xi_a} \pi (r_+^2 + a^2) +
\fft{4\pi\sigma_0^2}{\Xi_a} \Big(\alpha (1+ g_0^2 r_+^2) - 6\gamma g_0^2 (r_+^2 + a^2)\Big),\nn\\
J &=&-\fft{\partial G}{\partial \Omega}\Big|_{T}=\frac{a \left(r_+^2+a^2\right) \left(1+g_0^2 r_+^2\right)}{2\Xi_a^2 r_+}\Big(
\sigma_0  +4 g_0^2 \sigma _0^2 (\alpha -6 \gamma )\Big).
\eea
The mass, $M=G+T S + \Omega J$, turns out to be exactly $M=J/a$. The fact that the relation $J=Ma$ holds under the quadratic curvature correction is consistent with the fact that the correction in $D=4$ is trivial under field redefinition. This reproduces the special case of the more general four-derivative corrections to Einstein-Maxwell gravity \cite{Hu:2023gru}.

In five dimensions, there are two independent rotations, corresponding to two independent rotating parameters $(a_1,a_2)=(a,b)$. Specializing the general-$D$ formulae to $D=5$, we find that the corrected Gibbs free energy is
\bea
G_5 &=&  \frac{\sigma _0 \pi \left(r_+^2+a^2\right) \left(r_+^2+b^2\right) \left(1-g_0^2 r_+^2\right)}{8 \Xi_a \Xi_b\, r_+^2}-\frac{\pi  \sigma _0^2 \alpha\, \left(1+g_0^2 r_+^2\right){}^2}{4\Xi_a\Xi_b\, r_+^4 \left(r_+^2+a^2\right) \left(r_+^2+b^2\right)} \Big(9 r_+^8\nn\\
&& + 2 \left(a^2+b^2\right) r_+^6 + \left(a^4-20 a^2 b^2+b^4\right) r_+^4 -6 a^2 b^2 \left(a^2+b^2\right) r_+^2 +a^4 b^4\Big)\nn\\
&&+\frac{5\pi\sigma _0^2 g_0^2\,\gamma \left(r_+^2+a^2\right) \left(r_+^2+b^2\right)}{6 \Xi_a^2 \Xi_b^2\, r_+^2}\Big(g_0^2 r_+^2 (7 \Xi_a \Xi_b+\Xi_a+\Xi_b)-8 \Xi_a \Xi_b+\Xi_a+\Xi_b\Big).
\eea
This result recovers the work of \cite{Ma:2024ynp}, where its validity was verified by using a numerical perturbative method for the equal angular momentum case. The corrected Euclidean free energy was also later derived by the method of background subtraction \cite{Chen:2025ary}. It is understood that the temperature and two angular velocities are uncorrected, from which the remaining thermodynamic quantities can be easily obtained.

\subsubsection{All equal angular momenta}

When all the angular momenta are taken to be equal, $a_i=a$, the expression becomes significantly simpler. The corrected Gibbs free energy takes the form $ G=G_0+\delta G_{\alpha}+\delta G_{\gamma}$, where $G_0$ is the Gibbs free energy of the unperturbed rotating AdS black holes with all equal angular momenta
\bea
G_0=\left\{\begin{array}{lc}
		\frac{\sigma_0\mathcal{A}_{D-2}(r_+^2+a^2)^N(1-g_0^2 r_+^2)}{16\pi r_+^2\Xi_a^N }
\,,&\quad \hbox{$\epsilon=0$ ,} \\
&\\
		\frac{\sigma_0\mathcal{A}_{D-2}(r_+^2+a^2)^N(1-g_0^2 r_+^2)}{16\pi r_+\Xi_a^N }
\,,&\quad\hbox{$\epsilon=1$ .}\nn\\
	\end{array}\right.
\eea
The perturbative parts are given by
\bea
\delta G_{\alpha}&=&\left\{\begin{array}{lc}
	-\frac{\sigma_0^2\alpha\,(D-3)\mathcal{A}_{D-2}}{16\pi\Xi_a^N r_+^4(r_+^2+a^2)^{\frac{5-D}{2}}}(a^4-2(2D-3)a^2r_+^2+(D-2)^2r_+^4)\,,&\quad \hbox{$\epsilon=0$ ,}\\
&\\
	-\frac{\sigma_0\alpha\,(D-3)(D-2)\mathcal{A}_{D-2}}{16\pi\Xi_a^N r_+(r_+^2+a^2)^{\frac{6-D}{2}}}((D-2)r_+^2-2a^2)\,,&\quad\hbox{$\epsilon=1$ ,}
\end{array}\right.\nn\\
&&\nn\\
\delta G_{\gamma}&=&\left\{\begin{array}{lc}
	\frac{\sigma_0^2 \gamma\, D(D-1)\mathcal{A}_{D-2}g_0^2}{16 \pi (D-2) r_+^2}\frac{(r_+^2+a^2)^N}{\Xi_a^N}\Big[\frac{1}{2}(D^2-3D+4)g_0^2 r_+^2-
\frac{1}{2}(D^2-D-4)& \nn\\
	+(D-4)\frac{N}{\Xi_a}(1+g_0^2 r_+^2)\Big]\,, &\quad \hbox{$\epsilon=0$ ,}\\
&\\
	\frac{\sigma_0^2\gamma\, D(D-1)\mathcal{A}_{D-2}g_0^2}{16 \pi (D-2) r_+}\frac{(r_+^2+a^2)^N}{\Xi_a^{N}}\Big[(\frac{1}{2}D(D-2)(g_0^2r_+^2-1)&\nn\\
	+\frac{(D-4)(D-2)}{2}\frac{(D-2)(1+\Xi_a)}{2\Xi}(1+g_0^2 r_+^2) \Big]\,, &\quad\hbox{$\epsilon=1$ .}
\end{array}\right.
\eea
Since the temperature and angular velocities are unperturbed, given in \eqref{unperturbedthermo}, we can easily obtain the remaining thermodynamic quantities.

\section{Conclusion}

In this paper, we adopted the improved RS method and obtained the quadratic curvature correction to the on-shell Euclidean action of the rotating AdS black holes in general dimensions. A great simplification is achieved by the observation that the dependence of the cosmological constant in the Weyl-squared term of the rotating AdS black holes drops out completely, which enables us to use the corresponding result \cite{Wu:2024iiz} of the MP black holes. We then followed the procedure of field redefinition, developed for spherically-symmetric and static AdS black holes \cite{Hu:2023gru}, and obtained the general quadratic corrections. We obtained the corrected Gibbs free energy, associated with the thermodynamic ensemble where the temperature and angular velocities remain uncorrected. The general quadratic curvature corrections to all thermodynamic quantities of the rotating AdS black holes follow straightforwardly.

For the detractors, the method is predicated upon the assumption that the perturbed black hole solutions actually exist. For spherically-symmetric and static black holes, the existence of perturbed solutions can be easily verified \cite{Hu:2023gru}. It was also verified for asymptotically locally flat Taub-NUT black holes \cite{Chen:2024knw}. The testing of the RS method with perturbative solution is more involved for rotating black holes \cite{Cano:2019ore,Ma:2024ulp,Ma:2025vjk}, since there are no exact solutions, although for equal angular momenta, the equations simplify significantly in odd dimensions \cite{Ma:2020xwi,Mao:2023qxq}. The upshot is that there has so far been no counterexample for either asymptotically flat or AdS black holes in sphere topologies.

It can also be argued that there are only limited applications with the corrected thermodynamics, as the perturbative solutions are more useful. For the rotating cases, the higher-curvature perturbed solutions can be analytically obtained in terms of an additional power series expansion of small rotations \cite{Cano:2019ore,Ma:2024ulp,Ma:2025vjk}. However, these solutions are extremely complicated, and the RS method can provide a useful independent check whether the perturbed solutions are valid or not. Thus, our results pose both a challenge and independent verification for constructing the explicit perturbative solutions under the quadratic curvature extension.

\section*{Acknowledgement}

We are grateful to Liang Ma for useful discussions. This work is supported in part by the National Natural Science Foundation of China (NSFC) grants No.~12375052 and No.~11935009, and also by the Tianjin University Self-Innovation Fund Extreme Basic Research Project Grant No.~2025XJ21-0007.


\begin{thebibliography}{99}


%\cite{Stelle:1976gc}
\bibitem{Stelle:1976gc}
K.S.~Stelle,
``Renormalization of higher derivative quantum gravity,''
Phys. Rev. D \textbf{16}, 953-969 (1977)
doi:10.1103/PhysRevD.16.953
%2876 citations counted in INSPIRE as of 10 Feb 2026	
	
%\cite{Stelle:1977ry}
\bibitem{Stelle:1977ry}
K.S.~Stelle,
``Classical gravity with higher derivatives,''
Gen. Rel. Grav. \textbf{9}, 353-371 (1978)
doi:10.1007/BF00760427
%1360 citations counted in INSPIRE as of 10 Feb 2026

%\cite{Reall:2019sah}
\bibitem{Reall:2019sah}
H.S.~Reall and J.E.~Santos,
``Higher derivative corrections to Kerr black hole thermodynamics,''
JHEP \textbf{04}, 021 (2019)
doi:10.1007/JHEP04(2019)021
[arXiv:1901.11535 [hep-th]].
%99 citations counted in INSPIRE as of 09 Feb 2026

%\cite{Gibbons:1976ue}
\bibitem{Gibbons:1976ue}
G.W.~Gibbons and S.W.~Hawking,
``Action integrals and partition functions in quantum gravity,''
Phys. Rev. D \textbf{15}, 2752-2756 (1977)
doi:10.1103/PhysRevD.15.2752
%3875 citations counted in INSPIRE as of 09 Feb 2026

%\cite{York:1972sj}
\bibitem{York:1972sj}
J.W.~York, Jr.,
``Role of conformal three geometry in the dynamics of gravitation,''
Phys. Rev. Lett. \textbf{28}, 1082-1085 (1972)
doi:10.1103/PhysRevLett.28.1082
%1387 citations counted in INSPIRE as of 10 Feb 2026

%\cite{Wu:2024iiz}
\bibitem{Wu:2024iiz}
P.Y.~Wu and H.~L\"u,
``Quadratic curvature correction and its breakdown to thermodynamics of rotating black holes,''
Phys. Rev. D \textbf{111}, no.10, 104026 (2025)
doi:10.1103/ PhysRevD.111.104026
[arXiv:2405.04576 [hep-th]].
%3 citations counted in INSPIRE as of 09 Feb 2026

%\cite{Myers:1986un}
\bibitem{Myers:1986un}
R.C.~Myers and M.J.~Perry,
``Black holes in higher-dimensional space-times,''
Annals Phys. \textbf{172}, 304 (1986)
doi:10.1016/0003-4916(86)90186-7
%2197 citations counted in INSPIRE as of 09 Feb 2026

%\cite{Ma:2023qqj}
\bibitem{Ma:2023qqj}
L.~Ma, Y.~Pang and H.~L\"u,
``Higher derivative contributions to black hole thermodynamics at NNLO,''
JHEP \textbf{06}, 087 (2023)
[erratum: JHEP \textbf{08}, 118 (2024)]
doi:10.1007/ JHEP06(2023)087
[arXiv:2304.08527 [hep-th]].
%32 citations counted in INSPIRE as of 09 Feb 2026

%\cite{Lu:2025eub}
\bibitem{Lu:2025eub}
G.Y.~Lu, M.N.~Yang and H.~L\"u, ``Black hole mass/charge relation and weak no-hair theorem conjecture,'' JHEP \textbf{11}, 066 (2025)
doi:10.1007/JHEP11(2025)066
[arXiv:2508.14158 [hep-th]].
%3 citations counted in INSPIRE as of 23 Dec 2025

%\cite{Yang:2025rud}
\bibitem{Yang:2025rud}
M.N.~Yang, G.Y.~Lu and H.~L\"u,
``Black hole thermodynamics without black hole solutions,''
[arXiv:2512.09930 [hep-th]].
%0 citations counted in INSPIRE as of 23 Dec 2025

%\cite{Hawking:1982dh}
\bibitem{Hawking:1982dh}
S.W.~Hawking and D.N.~Page,
``Thermodynamics of black holes in anti-De Sitter Space,''
Commun. Math. Phys. \textbf{87}, 577 (1983)
doi:10.1007/BF01208266
%3019 citations counted in INSPIRE as of 09 Feb 2026

%\cite{Xiao:2023two}
\bibitem{Xiao:2023two}
Y.~Xiao and Y.Y.~Liu,
``First order corrections to black hole thermodynamics: A simple approach enhanced,''
Phys. Rev. D \textbf{110}, no.10, 104043 (2024)
doi:10.1103/Phys RevD.110.104043
[arXiv:2312.07127 [gr-qc]].
%16 citations counted in INSPIRE as of 09 Feb 2026

%\cite{Xiao:2025icr}
\bibitem{Xiao:2025icr}
Y.~Xiao and A.~Zhang,
``Validity of the background subtraction method for black hole thermodynamics in matter-coupled gravity theories,''
[arXiv:2511.07209 [gr-qc]].
%2 citations counted in INSPIRE as of 09 Feb 2026

%\cite{Guo:2025ohn}
\bibitem{Guo:2025ohn}
W.~Guo, X.~Guo, X.~Lan, H.~Zhang and W.~Zhang,
``Background subtraction method is not only much simpler, but also as applicable as the covariant counterterm method,''
Phys. Rev. D \textbf{111}, no.8, 084088 (2025)
doi:10.1103/PhysRevD.111.084088
[arXiv:2501.08214 [hep-th]].
%9 citations counted in INSPIRE as of 09 Feb 2026

%\cite{Chen:2025ary}
\bibitem{Chen:2025ary}
G.~Chen, X.~Guo, X.~Lan, H.~Zhang and W.~Zhang,
``Quadratic curvature corrections to five-dimensional Kerr-AdS black hole thermodynamics by the background subtraction method,''
Phys. Rev. D \textbf{112}, no.12, 124087 (2025)
doi:10.1103/kp7w-8st1
[arXiv:2508.18171 [hep-th]].
%4 citations counted in INSPIRE as of 09 Feb 2026

%\cite{Hawking:1998kw}
\bibitem{Hawking:1998kw}
S.W.~Hawking, C.J.~Hunter and M.~Taylor,
``Rotation and the AdS / CFT correspondence,''
Phys. Rev. D \textbf{59}, 064005 (1999)
doi:10.1103/PhysRevD.59.064005
[arXiv:hep-th/9811056 [hep-th]].
%655 citations counted in INSPIRE as of 10 Feb 2026

%\cite{Hu:2023gru}
\bibitem{Hu:2023gru}
P.J.~Hu, L.~Ma, H.~L{\"u} and Y.~Pang,
``Improved Reall-Santos method for AdS black holes in general 4-derivative gravities,''
Sci. China Phys. Mech. Astron. \textbf{67}, no.8, 280412 (2024)
doi:10.1007/s11433-024-2398-1
[arXiv:2312.11610 [hep-th]].
%17 citations counted in INSPIRE as of 09 Feb 2026

%\cite{Ma:2024ynp}
\bibitem{Ma:2024ynp}
L.~Ma, P.J.~Hu, Y.~Pang and H.~L\"u,
``Effectiveness of Weyl gravity in probing quantum corrections to AdS black holes,''
Phys. Rev. D \textbf{110}, no.2, L021901 (2024)
doi:10.1103/ PhysRevD.110.L021901
[arXiv:2403.12131 [hep-th]].
%23 citations counted in INSPIRE as of 09 Feb 2026

%\cite{Gibbons:2004uw}
\bibitem{Gibbons:2004uw}
G.W.~Gibbons, H.~L\"u, D.N.~Page and C.N.~Pope,
``The General Kerr-de Sitter metrics in all dimensions,''
J. Geom. Phys. \textbf{53}, 49-73 (2005)
doi:10.1016/j.geomphys.2004.05.001
[arXiv:hep-th/0404008 [hep-th]].
%470 citations counted in INSPIRE as of 09 Feb 2026

%\cite{Gibbons:2004js}
\bibitem{Gibbons:2004js}
G.W.~Gibbons, H.~L\"u, D.N.~Page and C.N.~Pope,
``Rotating black holes in higher dimensions with a cosmological constant,''
Phys. Rev. Lett. \textbf{93}, 171102 (2004)
doi:10.1103/PhysRevLett.93.171102
[arXiv:hep-th/0409155 [hep-th]].
%336 citations counted in INSPIRE as of 09 Feb 2026

%\cite{Balasubramanian:1999re}
\bibitem{Balasubramanian:1999re}
V.~Balasubramanian and P.~Kraus,
``A stress tensor for anti-de Sitter gravity,''
Commun. Math. Phys. \textbf{208}, 413-428 (1999)
doi:10.1007/s002200050764
[arXiv:hep-th/9902121 [hep-th]].
%2115 citations counted in INSPIRE as of 10 Feb 2026

%\cite{Bianchi:2001kw}
\bibitem{Bianchi:2001kw}
M.~Bianchi, D.Z.~Freedman and K.~Skenderis,
``Holographic renormalization,''
Nucl. Phys. B \textbf{631}, 159-194 (2002)
doi:10.1016/S0550-3213(02)00179-7
[arXiv:hep-th/0112119 [hep-th]].
%775 citations counted in INSPIRE as of 10 Feb 2026

%\cite{Brihaye:2008xu}
\bibitem{Brihaye:2008xu}
Y.~Brihaye and E.~Radu,
``Black objects in the Einstein-Gauss-Bonnet theory with negative cosmological constant and the boundary counterterm method,''
JHEP \textbf{09}, 006 (2008)
doi:10.1088/1126-6708/2008/09/006
[arXiv:0806.1396 [gr-qc]].
%80 citations counted in INSPIRE as of 10 Feb 2026

%\cite{Liu:2008zf}
\bibitem{Liu:2008zf}
J.T.~Liu and W.A.~Sabra,
``Hamilton-Jacobi counterterms for Einstein-Gauss-Bonnet Gravity,''
Class. Quant. Grav. \textbf{27}, 175014 (2010)
doi:10.1088/0264-9381/27/17/175014
[arXiv:0807.1256 [hep-th]].
%55 citations counted in INSPIRE as of 10 Feb 2026

%\cite{Cremonini:2009ih}
\bibitem{Cremonini:2009ih}
S.~Cremonini, J.T.~Liu and P.~Szepietowski,
``Higher derivative corrections to R-charged Black Holes: boundary counterterms and the mass-charge Relation,''
JHEP \textbf{03}, 042 (2010)
doi:10.1007/JHEP03(2010)042
[arXiv:0910.5159 [hep-th]].
%55 citations counted in INSPIRE as of 09 Feb 2026

%\cite{Lu:2011zk}
\bibitem{Lu:2011zk}
H.~L\"u and C.N.~Pope, ``Critical gravity in four dimensions,''
Phys. Rev. Lett. \textbf{106}, 181302 (2011)
doi:10.1103/PhysRevLett.106.181302
[arXiv:1101.1971 [hep-th]].
%265 citations counted in INSPIRE as of 09 Feb 2026

%\cite{Gibbons:2004ai}
\bibitem{Gibbons:2004ai}
G.W.~Gibbons, M.J.~Perry and C.N.~Pope,
``The first law of thermodynamics for Kerr-anti-de Sitter black holes,''
Class. Quant. Grav. \textbf{22}, 1503-1526 (2005)
doi:10.1088/0264-9381/22/9/002
[arXiv:hep-th/0408217 [hep-th]].
%512 citations counted in INSPIRE as of 09 Feb 2026

%\cite{Ma:2026cej}
\bibitem{Ma:2026cej}
L.~Ma, ``Black hole thermodynamic ensembles, Euclidean action and legendre transformation,'' [arXiv:2602.04954 [hep-th]].
%0 citations counted in INSPIRE as of 10 Feb 2026

%\cite{Chen:2024knw}
\bibitem{Chen:2024knw}
Y.Q.~Chen, H.S.~Liu and H.~L\"u,
``Taub-NUT black hole in higher derivative gravity and its thermodynamics,''
Phys. Rev. D \textbf{110}, no.10, 104068 (2024)
doi:10.1103/PhysRevD. 110.104068
[arXiv:2409.07692 [gr-qc]].
%0 citations counted in INSPIRE as of 12 Jan 2025

%\cite{Cano:2019ore}
\bibitem{Cano:2019ore}
P.A.~Cano and A.~Ruip{\'e}rez,
``Leading higher-derivative corrections to Kerr geometry,''
JHEP \textbf{05} (2019), 189
[erratum: JHEP \textbf{03} (2020), 187]
doi:10.1007/JHEP05(2019)189
[arXiv:1901.01315 [gr-qc]].
%152 citations counted in INSPIRE as of 11 Dec 2025

%\cite{Ma:2024ulp}
\bibitem{Ma:2024ulp}
L.~Ma, Y.~Pang and H.~L\"u,
``Leading higher derivative corrections to multipole moments of Kerr-Newman black hole,''
JHEP \textbf{02}, 079 (2025)
doi:10.1007/JHEP02(2025)079
[arXiv:2411.13639 [hep-th]].
%5 citations counted in INSPIRE as of 09 Feb 2026

%\cite{Ma:2025vjk}
\bibitem{Ma:2025vjk}
L.~Ma and H.~L\"u,
``Quadratic curvature correction to 5D Myers-Perry metric,''
[arXiv:2512.23797 [hep-th]].
%0 citations counted in INSPIRE as of 09 Feb 2026

%\cite{Ma:2020xwi}
\bibitem{Ma:2020xwi}
L.~Ma, Y.Z.~Li and H.~L\"u,
``$D = 5$ rotating black holes in Einstein-Gauss-Bonnet gravity: mass and angular momentum in extremality,''
JHEP \textbf{01}, 201 (2021)
doi:10.1007/ JHEP01(2021)201
[arXiv:2009.00015 [hep-th]].
%12 citations counted in INSPIRE as of 21 Apr 2024

%\cite{Mao:2023qxq}
\bibitem{Mao:2023qxq}
Q.Y.~Mao, L.~Ma and H.~L\"u,
``Horizon as a natural boundary,''
Phys. Rev. D \textbf{109}, no.8, 084053 (2024)
doi:10.1103/PhysRevD.109.084053
[arXiv:2307.14458 [hep-th]].
%2 citations counted in INSPIRE as of 06 May 2024

\end{thebibliography}
\end{document}